\documentclass[11pt]{article}

\usepackage{amsmath}
\usepackage{epsfig}

\makeatletter
\renewcommand\section{\@startsection {section}{1}{\z@}%
                                   {-3.5ex \@plus -1ex \@minus -.2ex}%
                                   {2.3ex \@plus.2ex}%
                                   {\normalfont\sf\large\bfseries}}
\renewcommand\subsection{\@startsection{subsection}{2}{\z@}%
                                     {-3.25ex\@plus -1ex \@minus -.2ex}%
                                     {1.5ex \@plus .2ex}%
                                     {\normalfont\sf\normalsize\bfseries}}
\renewcommand\subsubsection{\@startsection{subsubsection}{3}{\z@}%
                                     {-3.25ex\@plus -1ex \@minus -.2ex}%
                                     {1.5ex \@plus .2ex}%
                                     {\normalfont\sf\normalsize\bfseries}}
\renewcommand\paragraph{\@startsection{paragraph}{4}{\z@}%
                                    {3.25ex \@plus1ex \@minus.2ex}%
                                    {-1em}%
                                    {\normalfont\sf\normalsize\bfseries}}
\makeatother

\newcommand\DGKP{DGKP}
\newcommand\NNZ{NNZ(1994)}
\newcommand\NNPZ{NNPZ(1997)}

\newcommand\Ael{A_{\mbox{\footnotesize el}}}
\newcommand\calAel{{\mathcal A}_{\mbox{\footnotesize el}}}
\newcommand\tAel{{\tilde{A}}_{\mbox{\footnotesize el}}}

\newcommand\aem{\alpha_{\mbox{\footnotesize em}}}
\newcommand\as{\alpha_s}

\newcommand\calN{{\mathcal N}}

\newcommand\be{\begin{equation}}
\newcommand\ee{\end{equation}}

\def\be{\begin{equation}}
\def\ee{\end{equation}}
\def\bea{\begin{eqnarray}}
\def\eea{\end{eqnarray}}

\begin{document}
\begin{center}
 \LARGE \sf  {Impact parameter dependent $S$-matrix for dipole-proton scattering
 from diffractive meson electroproduction}{\renewcommand{\thefootnote}{\fnsymbol{footnote}}\footnote{
This research was partially supported by the EU Framework TMR
 programme, contract FMRX-CT98-0194, by the Polish Committee for Scientific Research     
grants Nos. KBN 2P03B 120 19, 2P03B 051 19, 5P03B 144 20 and by the US Department of
 Energy.}}
\end{center}
\vspace*{0.5cm}
\begin{center}
{\sc  S. Munier$^{(a)}$, A.M.~Sta\'sto$^{(a,b)}$, A.H.~Mueller$^{(c,d)}$}
\end{center}
\begin{center}
{\it
$^{(a)}$ INFN Sezione di Firenze, Largo E. Fermi 2, 50125 Firenze, Italy. \\
$^{(b)}$ Department of Theoretical Physics, H. Niewodnicza\'nski Institute
of Nuclear Physics, 31-341 Krak\'ow, Poland. \\
$^{(c)}$ Department of Physics, Columbia University, New York,
N.Y. 10027.\\
$^{(d)}$ Laboratoire de Physique Th\'eorique, Universit\'e Paris-Sud, F-91405
Orsay Cedex, France.}\\
\end{center}
\begin{abstract}
We extract the  $S$-matrix element for dipole-proton
scattering using the data
on diffractive electroproduction of vector mesons at HERA. 
By considering the full $t$ dependence of this process we are able to 
reliably unfold the profile of the $S$-matrix 
for impact parameter values $b > 0.3 \, \rm fm$ .
We show that the results depend only weakly on the  choice of the
form for the vector meson wave function. We relate this result
to the discussion about possible saturation effects at HERA.
\end{abstract}

\section*{Introduction}

One of the key issues in deeply inelastic scattering and deeply inelastic
diffraction is whether parton saturation effects are present in the HERA
energy regime. The answer to this question has important implications for
our understanding of the very early stages of relativistic heavy ion
collisions where parton saturation would result in the production of a
high density and high field strength, $F_{\mu\nu}\sim
1/{\sqrt{\alpha_s}}\ ,$ state of QCD.  Such a state would be a new and
exciting regime of nonperturbative but small coupling QCD.\\

Parton saturation can be viewed as a manifestation of unitarity limits
being reached.  However, unitarity limits are difficult to see directly
in deep inelastic scattering and diffraction.  Thus, most studies of
parton saturation at HERA have been done in terms of models which
explicitly impose unitarity at moderate $Q^2$ and small $x$.  The most
widely used such model is that of Golec-Biernat and W\"usthoff
\cite{GBW} where the
$S$-matrix for the scattering of a dipole of separation $r$   on
a proton is given by $S = e^{-r^2Q_s^2/4}$ with the saturation momentum,
$Q_s^2,$ depending on the energy of the scattering.  The fact that the
Golec-Biernat W\"usthoff model and its generalization work so well at
HERA is perhaps some evidence that deep inelastic scattering, and
especially diffraction, have reached unitarity limits.\\

It would be nice to see the attainment of unitarity limits directly
without having to use a model.  This is not an easy task.  Recall that in
proton-proton scattering the energy dependence of total and elastic cross
sections changes little between ISR and Tevatron energies.  The near
saturation of unitarity limits was found in proton-proton scattering
only when Amaldi and Schubert \cite{AMALDI} evaluated the proton-proton elastic
scattering amplitude in impact parameter space and found near blackness
for small impact parameters.\\

Vector meson production seems to be the best process from which to
extract the dipole proton elastic scattering amplitude.  For example,
it is easy to check that our procedure 
(see Eqs.~(\ref{eq:dsigmadt})-(\ref{eq:means}) below) would not
work for the total diffractive cross section because there is not a single
definite state produced, a condition which seems necessary in the way we
proceed.\\

In this paper we copy the Amaldi-Schubert analysis in order to determine
the dipole-proton scattering amplitude in impact parameter space in 
terms of diffractive $\rho-$production data at HERA.  Recall that
diffractive electroproduction of a $\rho-$meson can be viewed in terms
of the following sequence of transitions:  (i)  The virtual photon goes
into a quark-antiquark pair (dipole).  (ii)  The dipole  scatters
elastically on the proton.  (iii)  The dipole then goes into a
$\rho-$meson.  Thus $\rho-$electroproduction naturally carries
information on the dipole-quark elastic scattering amplitude.  The main
uncertainty is in the wavefunction of the $\rho-$meson which appears to
be reasonably well constrained by previous phenomenology.\\

The dipole-proton $S$-matrix which we extract as a function of impact
parameter is averaged over the dipole sizes which naturally appear in the
$\gamma^\ast$ and $\rho-$meson wavefunctions.  This averaging does not
affect the fact that for weak interactions $S$\ is near 1 and for
strong interactions  $S$\  is near zero.  Thus the smallness of   $S$\
is a measure of how close one is to the unitarity limit, $S=0.$  Perhaps
the best way to measure how close the dipole-proton scattering is to
the unitarity limit is to note that $\langle 1-S^2\rangle \ge
1-\langle S\rangle^2$ gives the
probability that a dipole passing the proton will induce an inelastic
reaction at the impact parameter in question  ($\langle\cdot\cdot\cdot\rangle$
denotes the average over dipole sizes as explained in Eq.~(\ref{eq:mean})).  From our
analysis this probability is likely significantly greater than $1/2$ for
$Q^2 \leq 2\,\mbox{GeV}^2$ and for $b$ near zero (see Fig.~\ref{fig:results}), although better
large momentum transfer data would be needed to definitively determine  
$S$ at $b\approx 0.$  Thus, it is likely that saturation effects are
important in this low $Q^2$ regime.  Our estimate of $Q_s^2$ is about
$1-1.5\,\mbox{GeV}^2$ at $ b\approx 0.3$ fm, although this estimate has large
uncertainties because it depends on knowing the average dipole size as a
function of $Q^2.$  Finally, we determine the total dipole-proton cross
sections for $x\approx 10^{-4}$ to be 14.9, 10.6 and 7.5 mb at $Q^2=0.45, 3.5$ and $7\,\mbox{GeV}^2$
respectively.\\

The outline of the paper is as  follows:
in section~\ref{sec:1} we establish  the relation between the $S$-matrix for
dipole-proton scattering averaged over the distribution of dipole
sizes and the diffractive differential cross section for
electroproduction of  vector mesons.
 Due to the particular properties of this distribution, this formula translates
into a method of determining the $S$-matrix profile in impact 
parameter space for a dipole of given size $r$.
The only theoretical input needed is the vector meson wave function.
In section~\ref{sec:2} we discuss the parametrizations for the wave functions that we
used, and
we study in detail the model independence of our procedure.
In section~\ref{sec:3} we apply this method to
the analysis of the available HERA data and present our main results
on the $S$-matrix profile in impact parameter space.  We discuss the uncertainties
due to the experimental errors on the measured cross section and due to
the lack of data in the high $t$ region.
We also discuss the possible implications for studies of saturation
effects at high energies.
Finally, in section~\ref{sec:4} we state the summary of our results.
We append some more technical points.

\section{Dipole-proton $S$-matrix and diffractive meson production cross section}

\label{sec:1}
\begin{figure}[ht]
\begin{center}
\epsfig{file=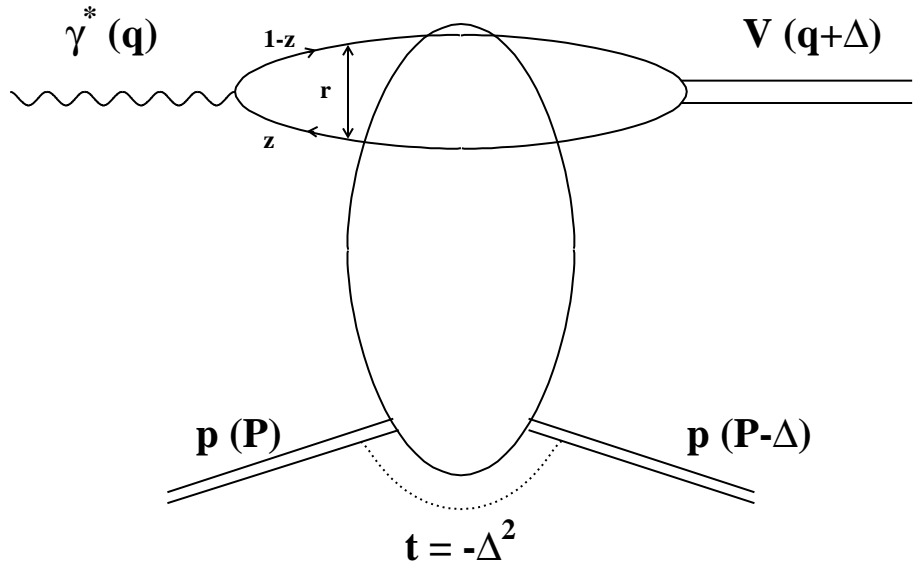,width=15cm}
\end{center}
\caption{Diagrammatic representation of diffractive production
of vector mesons. $\gamma^*(q)$ is the virtual photon of momentum $q$,
$V (q+\Delta)$ is the produced vector meson with momentum $q+\Delta$.
The target proton is scattered elastically.}
\label{fig:diagram}
\end{figure}

We consider the process of  diffractive production of a vector meson,
$\gamma^*\!-\!p\rightarrow V\!-\!p$, shown in Fig.~\ref{fig:diagram}. 
We adopt here the dipole formulation \cite{DIPOLENIK,DIPOLEMUELL} in which
the virtual photon $\gamma^*$ fluctuates into a $q\bar{q}$ pair (dipole) 
which subsequently  interacts elastically with a proton, and eventually forms a
meson bound state.
This picture is justified at high center-of-mass energy of
the $\gamma^* p$ system, since in this regime it is guaranteed that
these successive processes are clearly separated in time.
We assume that the energy is sufficiently high
so that $s$-channel helicity conservation holds to a good accuracy,
and we limit ourselves to the case of longitudinal vector meson production.
The kinematics of this process is shown in  Fig.~\ref{fig:diagram}:
we define the 2-momentum transfer ${\mathbf \Delta}$ related to the Mandelstam 
variable $t$ by $t = -{\mathbf \Delta}^2$. $Q^2=-q^2$ and
$x=Q^2/(2P\cdot q)$ are as
usual the virtuality of the photon and the Bjorken variable,
respectively. 

In this framework, the amplitude
$\calAel(x,{\mathbf \Delta},Q)$ for the scattering can be written 
in the following factorized form:
\begin{equation}
 \calAel(x,{\mathbf \Delta},Q) = 
\sum_{h,\bar h}\int d^2\mathbf r\,dz\,\psi_{\gamma^*}^{h,\bar
h *}(z,{\mathbf r};Q)\,\Ael^{q\bar{q}-p}(x,{\mathbf r},{\mathbf \Delta})\,
\psi^{h,\bar h}_V(z,{\mathbf r}) \; ,
\label{eq:master}
\end{equation}
where $\Ael^{q\bar{q}-p}(x,{\mathbf r},{\mathbf \Delta})$ is the
elementary amplitude 
for the elastic scattering of a dipole of size ${\mathbf r}$ on the
proton. $\psi^{h,\bar h}_{\gamma^*}$ is the photon light-cone wave function
projected onto a state made of
a quark-antiquark pair of charges $e_q$, masses $m_q$ and respective helicities $h$
and $\bar h$. This function is computed in first order light-cone
perturbation theory of quantum electrodynamics, and for 
the longitudinally polarized virtual photon, it reads 
\cite{DIPOLENIK,Bjorken:1971ah}
\begin{equation}
\psi^{h,\bar{h}}_{\gamma^*}(z,{\mathbf r};Q)
=\delta_{h,-\bar h} \ { e_q \over \sqrt{4 \pi}} \ \frac{\sqrt{N_c}}{2\pi}
2z(1\!-\!z)Q\,K_0(\varepsilon r)\ ,
\label{eq:photon}
\end{equation}
where
\begin{equation}
\varepsilon^2 =  Q^2 z(1-z) +  m_q^2 \ .
\end{equation}
Similarly, $\psi^{h,\bar h}_V$ is the  wave function of the vector meson produced in the final state.
We discuss it in more detail in the next section.
The variable $z$ in Eq.~(\ref{eq:master})
is the fraction of longitudinal momentum of the virtual photon carried
by the quark. 

The amplitude is normalized in such a
way that the differential cross section for the full process is
\begin{equation}
{d\sigma \over dt}  = {1 \over 16 \pi} | \calAel(x,{\mathbf \Delta},Q) |^2 \; .
\label{eq:dsigmadt}
\end{equation}

Furthermore, the elementary amplitude $\Ael^{q\bar{q}-p}(x,{\mathbf
 r},{\mathbf \Delta})$ can be related to 
the $S$-matrix element $S(x,{\mathbf r},{\mathbf b})$ for the scattering of a
dipole of size ${\mathbf r}$ at impact parameter ${\mathbf b}$ \cite{MUELLER}
\begin{equation}
\Ael^{q\bar{q}-p}(x,{\mathbf r},{\mathbf \Delta})
=\int  d^2{\mathbf b}\,
{\tAel}^{q\bar{q}-p}(x,{\mathbf r},{\mathbf b})
e^{i\mathbf{b} {\mathbf \Delta}}
 =2\int d^2{\mathbf b}\,[1-S(x,{\mathbf
r},\mathbf{b})]e^{i\mathbf{b} {\mathbf \Delta}}\ .
\label{eq:s}
\end{equation}
We can check briefly the consistency of this formula, 
which defines $S(x,{\mathbf r},{\mathbf b})$ (see Ref.~\cite{landau}).
In the standard normalizations adopted here for the amplitudes, 
the optical theorem takes the form 
$\sigma_{\mbox{\footnotesize tot}}^{q\bar q-p}(x,{\mathbf r})
={\mathcal I}m\, i\Ael^{q\bar q-p}(x,{\mathbf r},{\mathbf\Delta}\!=\!0)$.
Using formula~(\ref{eq:s}), 
this relation translates into the following expression
for the total dipole-proton cross section:
\begin{equation}
\sigma_{\footnotesize\rm tot}^{q\bar q-p}(x,\mathbf r)=2\int d^2{\mathbf
b}\,(1-{\mathcal R}e\,S(x,\mathbf r,\mathbf b))\ .
\label{eq:landau}
\end{equation}
On the other hand, taking the amplitude~(\ref{eq:s}) squared, dividing it
by the flux factor and integrating it over phase space, it is easy to see that the 
elastic cross section is given by
\begin{equation}
\sigma_{\footnotesize\rm el}^{q\bar q-p}(x,\mathbf r)=
\phantom{2}\int d^2{\mathbf b}\,\left|1-S(x,\mathbf r,\mathbf b)\right|^2\ .
\label{eq:landau2}
\end{equation}
When $S=0$, the unitarity limit is reached which is equivalent to the
scattering on a black body. One sees from
formulae~(\ref{eq:landau}) and~(\ref{eq:landau2}) that the elastic cross
section is half the total one, as it should be.
This justifies the consistency of the normalization adopted for 
$S(x,{\mathbf r},{\mathbf b})$ in Eq.~(\ref{eq:s}).
We will assume in the following that  $i\Ael^{q\bar q-p}(x,{\mathbf r},{\mathbf \Delta})$
is purely imaginary, i.e. that $S(x,{\mathbf r},{\mathbf b})$ is real.
\\

The main goal of this paper is to extract  $S(x,{\mathbf r},{\mathbf
b})$ taking advantage of
the present experimental knowledge of the cross section for diffractive vector meson
production, see Eq.~(\ref{eq:master}).
To this aim we express the amplitude which appears in
Eqs.~(\ref{eq:master},\ref{eq:dsigmadt}) 
by means of $S(x,{\mathbf r},{\mathbf b})$ using
 relation (\ref{eq:s}). Then we take the inverse Fourier transform 
of the square root of Eq.(\ref{eq:dsigmadt}), which gives
\begin{equation}
\int { d^2 {\mathbf \Delta} \over (2\pi)^2 }\, \sqrt{{d\sigma \over dt}}
 e^{-i 
{\mathbf \Delta} {\mathbf b}} \; = \; {1 \over \sqrt{16 \pi}} \, \sum_{h,\bar{h}} \;
\int d^2 {\mathbf r}\,  dz \, \psi_{\gamma^*}^{h,\bar{h} *}(z,{\mathbf r},Q) \, 2 \, [1-S(x,{\mathbf r},
{\mathbf b})]  \,\psi_V^{h,\bar{h}}(z,{\mathbf r}) \ .
\label{eq:master1}
\end{equation}
In the following we suppress the angular dependence of the $S$-matrix
since in this process one is only sensitive to the quantities
which are angular averaged.
We single out the (logarithmic) distribution of dipole sizes at the photon vertex,
which is the overlap between the photon
and the meson wave functions
\begin{equation}
p(r,Q)\equiv 2\pi\,r^2\,\sum_{h,\bar h}
\int_0^1 dz\,\psi^{h,\bar h*}_{\gamma^*}(z,r;Q)\,\psi^{h,\bar h}_V(z,r)\ ,
\label{eq:distrib}
\end{equation}
and we denote $N(Q)$ its integral
\begin{equation}
N(Q)\equiv\int_{0}^\infty {dr \over r} \,p(r,Q) \; .
\label{eq:intdistrib}
\end{equation}
We then define the mean value of a given
function $f(r)$ with respect 
to the probability measure $p(r,Q)/N(Q)$ by
\begin{equation}
\langle f(r) \rangle_p \; \equiv \; \int_0^{\infty} {dr \over r} {p(r,Q)  \over N(Q)} \, f(r) \, .
\label{eq:mean}
\end{equation}
We now rewrite Eq.~(\ref{eq:master1}) 
using these definitions~(\ref{eq:distrib}), (\ref{eq:intdistrib}),
(\ref{eq:mean}), and we obtain the following formula:
\begin{equation} 
\langle S(x,r,b)\rangle_p=1-\frac{1}{2 N(Q)\pi^{3/2}}
\int{d^2{\mathbf \Delta}}\,e^{-i{\mathbf
\Delta}{\mathbf b}}\sqrt{\frac{d\sigma}{dt}}\ ,
\label{eq:means}
\end{equation}
which shows that the measurement of the differential cross section $d\sigma/dt$
enables us to determine the $S$-matrix at fixed impact parameter $b$,
averaged over the dipole size distribution $p(r,Q)$ defined by Eq. (\ref{eq:distrib}). 

One should stress that $N(Q)$ which appears in Eq.~(\ref{eq:means})
is the only source of theoretical uncertainty in this formula for
the average $S$-matrix. This quantity depends on the parametrization of the wave function
of the vector meson. However, as we shall discuss in detail 
in  Sec.~\ref{sec:2}, $N(Q)$ is very
well constrained 
and is hardly model-dependent. We now have to investigate the meaning
of this average over dipole sizes, by studying $p(r,Q)$ in more detail.\\

\begin{figure}[ht]
\begin{center}
\epsfig{file=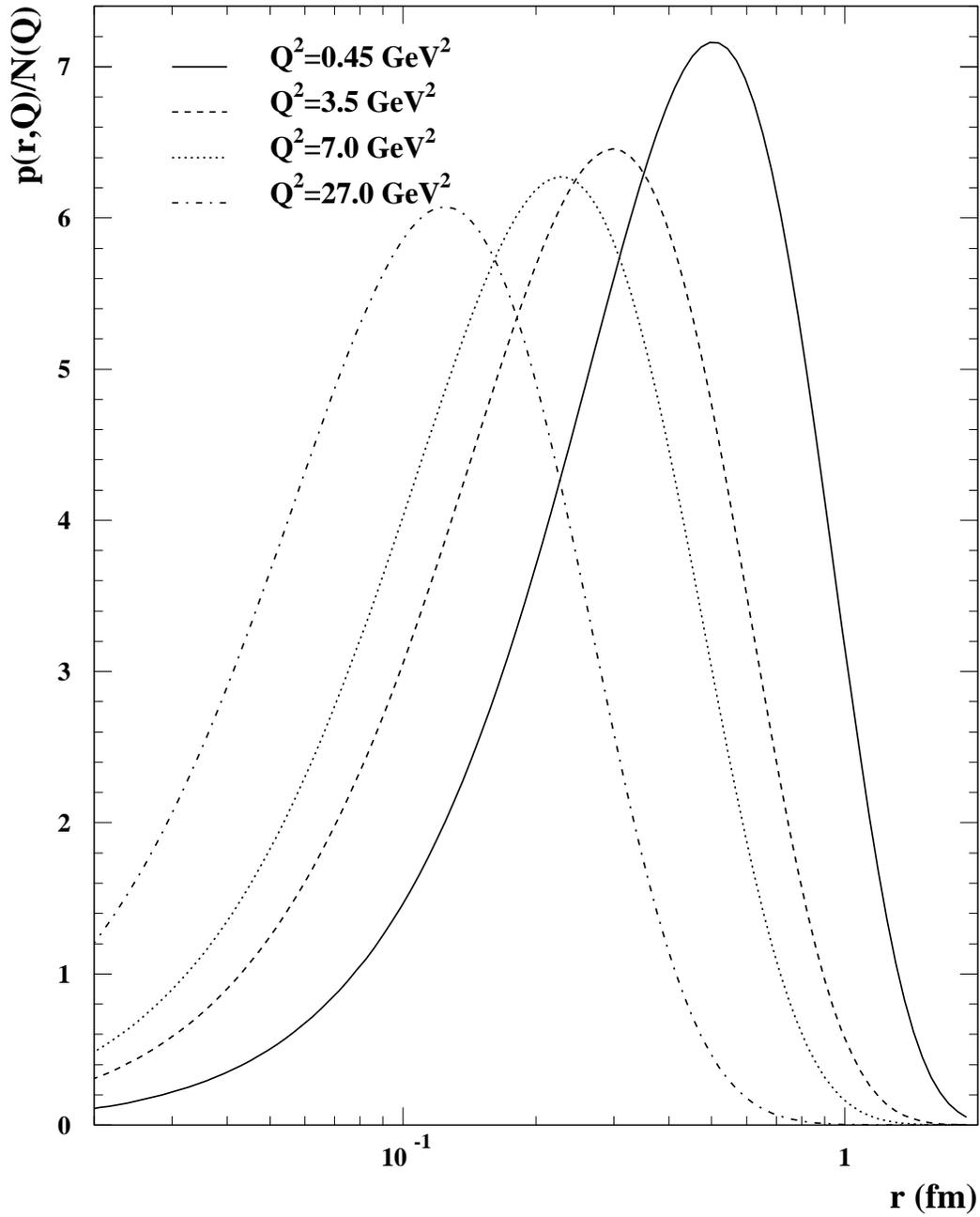,width=15cm}
\end{center}
\caption{Logarithmic distribution of dipole sizes at the photon
vertex $p(r,Q)/N(Q)$, for different $Q^2$.}
\label{fig:proba}
\end{figure}

The distribution $p(r,Q)$ has some interesting properties which
are quite general and rather independent of the particular model for 
the meson wave function: 
it is  sharply peaked at a specific value of $r$, which
depends on $Q^2$ like $r_Q = A/\sqrt{Q^2 + m_V^2}$
and its width, roughly independent of $Q^2$, is of order unity on 
a logarithmic scale.
To illustrate these properties, we choose a particular model 
for $\psi_V(z,r)$ \cite{DOSCH} (details will be given in the next
section), and we  plot $p(r,Q)$  for various
values of $Q^2$ in Fig.~\ref{fig:proba}.
One can easily obtain a rough estimate of $r_Q$ considering the fact that
on one hand, the photon wave function behaves like
$\log r$ at small $r$, and on the other hand, it
can be approximated by $\exp(-\varepsilon r)$
in the asymptotic region $r\rightarrow\infty$, see Eq.~(\ref{eq:photon}).
Assuming furthermore that the vector meson wave function is smooth in $r$, the integrand 
$p(r,Q)/r \sim r K_0(\varepsilon r)$ is then peaked around the maximum
of the function $r\exp(-\varepsilon r)$, i.e. around
$r_Q \sim 1/\varepsilon$. In the case of longitudinal 
vector meson production, the dominant contribution comes from
symmetric configurations for which
$z$ is close to $1/2$.
This leads to the above-quoted formula for $r_Q$, with $A \simeq 2$.

When $\langle S(x,r,b)\rangle_p$ is significantly below $1$, the quantity
$r_Q$ can be interpreted as the mean value of the sizes $r$ of the dipoles
which participate in the interaction. 
In this case, the dipole-proton 
amplitude
${\tAel}^{q\bar{q}-p}(x,{ r},{ b})$
is large for any dipole of size $r$ present
in the initial wave function 
and thus does not filter these dipoles:
the dominant dipole sizes involved in the interaction are then
effectively distributed
according to $p(r,Q)$. 
However, in the opposite case when $\langle S(x,r,b)\rangle_p \sim 1$, 
the amplitude ${\tAel}^{q\bar{q}-p}(x,{r},{ b})$ is small for the
relevant values of $r$.
In this situation, we are sensitive to the behaviour of the dipole cross section
at small $r$, i.e. ${\tAel}^{q\bar{q}-p}(x,{ r},{
b})\sim r^2$. This is known as the colour-transparency
property. Then
the interacting dipoles are effectively distributed according to
$r^2 p(r,Q)$. This results in the dominant dipole sizes being shifted
towards larger values\footnote{The quantity $r_s$
in this case is the so called ``scanning radius'' introduced in
\cite{NIKO2}.} of $r \simeq r_s$ for which an estimate can be provided
using the same technique as previously, and yields
$r_s \simeq 6/\sqrt{Q^2+m_V^2}$.

According to this discussion and considering the fact that one is {\it
a priori} more
interested in the case when 
$\langle S(x,r,b)\rangle<1$, one can make the following approximation:
\begin{equation}
\langle S(x,r,b) \rangle_p \simeq S(x,r_Q,b) \qquad \mbox{with} \qquad
r_Q \; \simeq \; { A \over\sqrt{ Q^2 + m_V^2}} \; , \quad A \simeq 2 \; .
\label{eq:rq}
\end{equation}
Thus for fixed $Q^2$ and energy Eqs.~(\ref{eq:means},\ref{eq:rq}) enable
us to determine $S(x,r_Q,b)$, which is the $S$-matrix element for the
scattering of a dipole of size $r_Q$ on a proton, at energy $Q^2/x$
and impact parameter $b$. 
A novel feature of our analysis is the fact that
we are able to extract the profile of the $S$-matrix (or equivalently
of the dipole-proton amplitude) in impact parameter space.
The previous studies of Refs.~\cite{GBW,NIKO2,NIKO1,MFGS,RUETER,SHAW,CVETIC,CAPELLA}  
provided an estimate of the dipole cross section integrated over the
impact parameter.


\section{From mesons to dipoles}
\label{sec:2}

As already mentioned earlier in Sec.1 the only theoretical uncertainty 
in the extraction of $S$-matrix profile in impact parameter space
is the form of the vector meson wave function which occurs
in the overlap function $p(r,Q)$ (Eq.~(\ref{eq:distrib})) and in its integral $N(Q)$ 
(Eq.~(\ref{eq:means})). 
In this section, we present the different models that we choose for
the vector meson wave function, then we compare them
phenomenologically, and finally we evaluate the uncertainty on
$S(x,r_Q,b)$ induced by the freedom of choice of the vector meson wave function.

\subsection{Models for the vector meson wave function}

Several different models
for the vector meson wave function exist in the literature
\cite{DOSCH,NIKO2,NIKO1,BRODSKY,FKS}.  
All these approaches use the information from spectroscopic
models of long distance physics. 
In these models one assumes that the meson is composed of a constituent
quark and antiquark which move in an harmonic oscillator potential.
This results in a wave function
which has a gaussian dependence on the spatial separation between the quarks.
Additionally these models are supplemented by the short-distance physics
driven by the QCD exchange of  hard gluons between the valence quarks
of the vector meson. Finally, a relativization procedure has to be
applied.

Of course, there are many  uncertainties in the above procedure of
obtaining the wave functions 
for the vector mesons, however they are  constrained
by model independent features.
First of all, $\psi_V^{h,\bar{h}}$ has to satisfy the following normalization
condition \cite{DOSCH}
\begin{equation}
1=\sum_{h,\bar h}{\int d^2{\mathbf r}\,dz\,|\psi^{h,\bar
h}_V(z,{\mathbf r})|^2}\ .
\label{eq:norm}
\end{equation}
Second, the value of the wave function at the origin is related
to the leptonic decay width $\Gamma(V\rightarrow e^+ e^-)$ of the vector meson
given by the following formula
\begin{equation}
\int_0^1 dz\,\psi_V(z,r=0)=\sqrt{\frac{\pi}{N_c}}\frac{f_V}{2\hat
e_V},\quad\mbox{where}\quad
\langle0|J^\mu_{\mbox{\footnotesize em}}(0)|V\rangle\equiv
e_q f_V\,m_V\,\varepsilon^\mu\ ,
\label{eq:lwid}
\end{equation}
where $f_V$ is the coupling of the meson to the electromagnetic current and 
$\hat e_V$ the isospin factor, which is the
effective charge of the quarks in units of the elementary charge $e$:
for the $\rho$ meson it is the charge of the combination 
$(u\bar u\!-\!d\bar d)/\sqrt{2}$, i.e. $\hat{e}_V = 1/\sqrt{2}$.
Finally, one requires that the mean radius 
be consistent with the electromagnetic radius of the vector meson.

In our calculation we consider the models proposed in
Ref.~\cite{DOSCH} (which we refer to as \DGKP)
and \cite{NIKO2,NIKO1} (which we denote \NNZ~and \NNPZ~respectively).

\paragraph{Model \DGKP} The following wave function is adopted:
\begin{equation}
\psi_{V}^{h,\bar{h}}(z,r) \; = \; \delta_{h,-\bar{h}} \, z(1-z) \, 
\frac{1}{\sqrt{4\pi}}
{\pi f_V \over \sqrt{N_c} \hat{e}_V } \, f(z) \, \exp(-\omega^2 r^2 /2) \; .
\label{eq:doschwf}
\end{equation}
The parameter  $\omega$ and the overall normalization
are fixed by the  condition
(\ref{eq:norm}) and by the value of the leptonic width.
(The exact values of these parameters as well as the form of the function
$f(z)$ are given in appendix A).
Using the above form of the wave function $\psi_V$ one can write
$p(r,Q)$ by replacing it in Eq.~(\ref{eq:distrib}). In the case of model \DGKP, it reads:
\begin{equation}
p(r,Q)
 =  2\pi r^2\int { dz \over 4 \pi}
f_V \; e_q \; z(1-z)f(z)
e^{-\omega^2 r^2/2} \, 2 \, z (1-z) Q K_0(\varepsilon r) \; .
\label{eq:doschp}
\end{equation}

\paragraph{Models \NNZ~and \NNPZ}  The wave function of the vector meson
is given by 
\begin{equation}
\psi_V^{h,\bar{h}}(z,{\mathbf r})
 \; = \;  \delta_{h,-\bar{h}} \sqrt{N_c \over 4\pi }
{1 \over m_V z(1-z)} [m_V^2 z (1-z) - {\mathbf \nabla_r}^2 + m_q^2] \phi(r,z) \, .
\label{eq:nikowf}
\end{equation}
The form of the radial wave function $\phi(r,z)$ is given in appendix A.
The expression for the overlap $p(r,Q)$ then is
\begin{equation}
p(r,Q)
= 2\pi r^2 {N_c \hat{e}_V \over (2\pi)^2}
{2 \, e_q \, Q \over m_V} \int dz \bigg\{  
[m_V^2 z (1-z) + m_q^2] \, 
K_0(\varepsilon r )- \epsilon K_1(\varepsilon r) {\partial \over \partial r} 
\bigg\} \phi(r,z) \; .
\label{eq:nikop}
\end{equation}
The models \NNZ~and \NNPZ~differ only by the tuning of the parameters.

Expressions (\ref{eq:doschp}) and (\ref{eq:nikop}) differ on several
points. We study the influence of these differences on the
distribution $p(r,Q)$ and on its integral $N(Q)$ in the next sections.


\subsection{Comparison of the theoretical predictions with the HERA data at $t=0$}

Before applying the models  (\ref{eq:doschwf},\ref{eq:doschp}) and 
(\ref{eq:nikowf},\ref{eq:nikop}) for the vector meson wave functions 
to our procedure of extracting the $S$-matrix element, we
additionally checked
how they compared with the data on forward diffractive vector meson
electroproduction at HERA.

To this aim we compute $d\sigma /dt |_{t=0}$
using formulae (\ref{eq:master},\ref{eq:dsigmadt}) and substitute
the dipole-proton amplitude $\Ael^{q\bar q-p}$ by the one proposed in 
Golec-Biernat and W\"usthoff's
saturation model\footnote{
A similar study was done in detail recently in \cite{CALDWELL} 
where different forms of
the wave functions for $\rho_0$ and $J/\Psi$ were considered and used
in the calculation of the forward longitudinal and transverse total cross sections.
The meson wave functions were chosen according to 
general principles rather than just relying on existing models.}
($\sigma(x,r)$ in the notations of Ref.~\cite{GBW}).

\begin{figure}[htb]   
\begin{center}
\epsfig{file=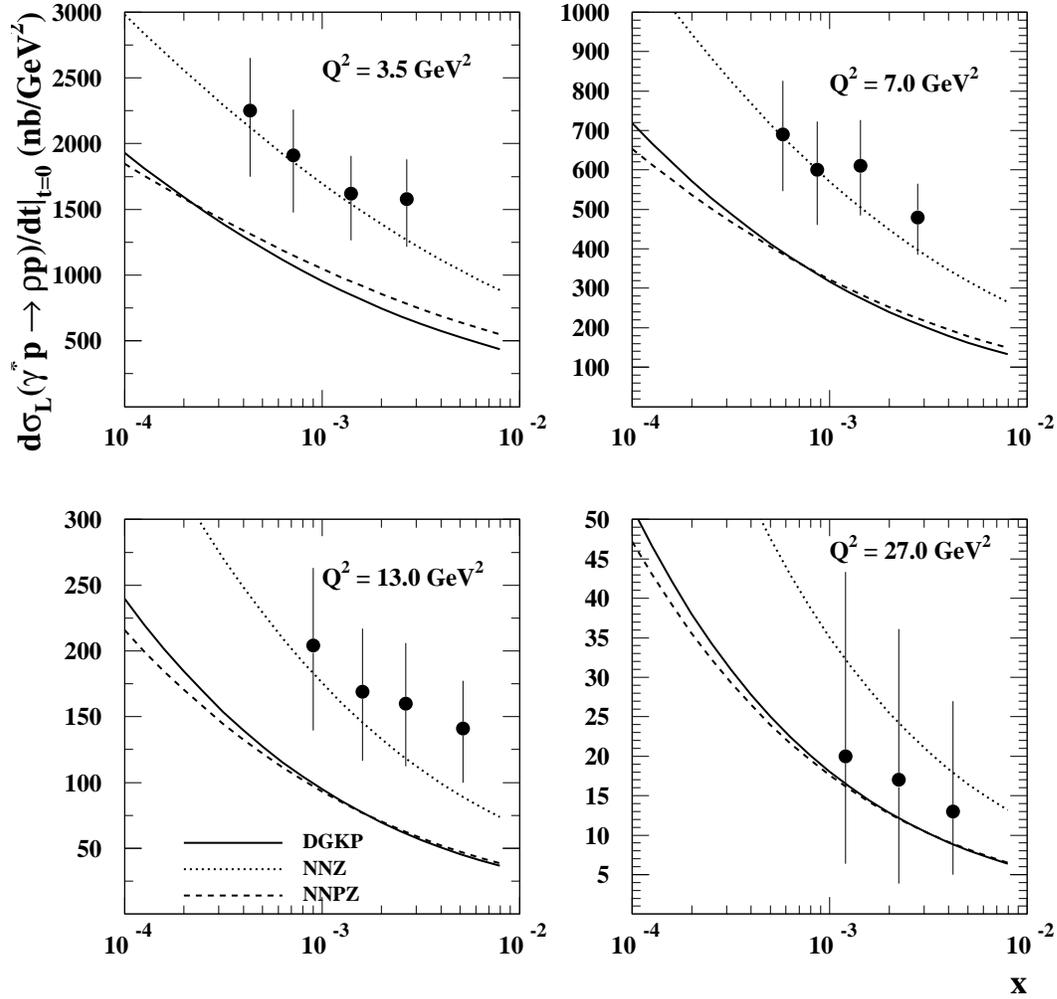,width=15cm}
\end{center}
\caption{The forward longitudinal cross section for the production of $\rho_0$
vector meson. Data are from \cite{ZEUS} and the three curves
correspond to different assumptions 
for the wave functions of the vector meson. Solid curve: model
\DGKP~\cite{DOSCH}, 
dashed curve: model \NNPZ~\cite{NIKO1} and dotted curve 
model \NNZ~\cite{NIKO2}. The dipole cross
section is taken from Golec-Biernat and W\"usthoff model \cite{GBW}. }
\label{fig:total}   
\end{figure}

The theoretical curves
corresponding to all three different models \DGKP, \NNZ, \NNPZ~are 
plotted in Fig.~\ref{fig:total}.
Let us stress that there is no additional tuning of any of the parameters
of the wave functions. 
One of the models, \NNZ, is able to describe the
data quite well, whereas the other ones underestimate the
normalization by as much as a factor of 2.
However, the energy dependence
of the cross section is well described by all models.
Qualitatively similar conclusions were reached in the study of Ref.~\cite{CALDWELL}.

\begin{figure}[htb]   
\begin{center}
\epsfig{file=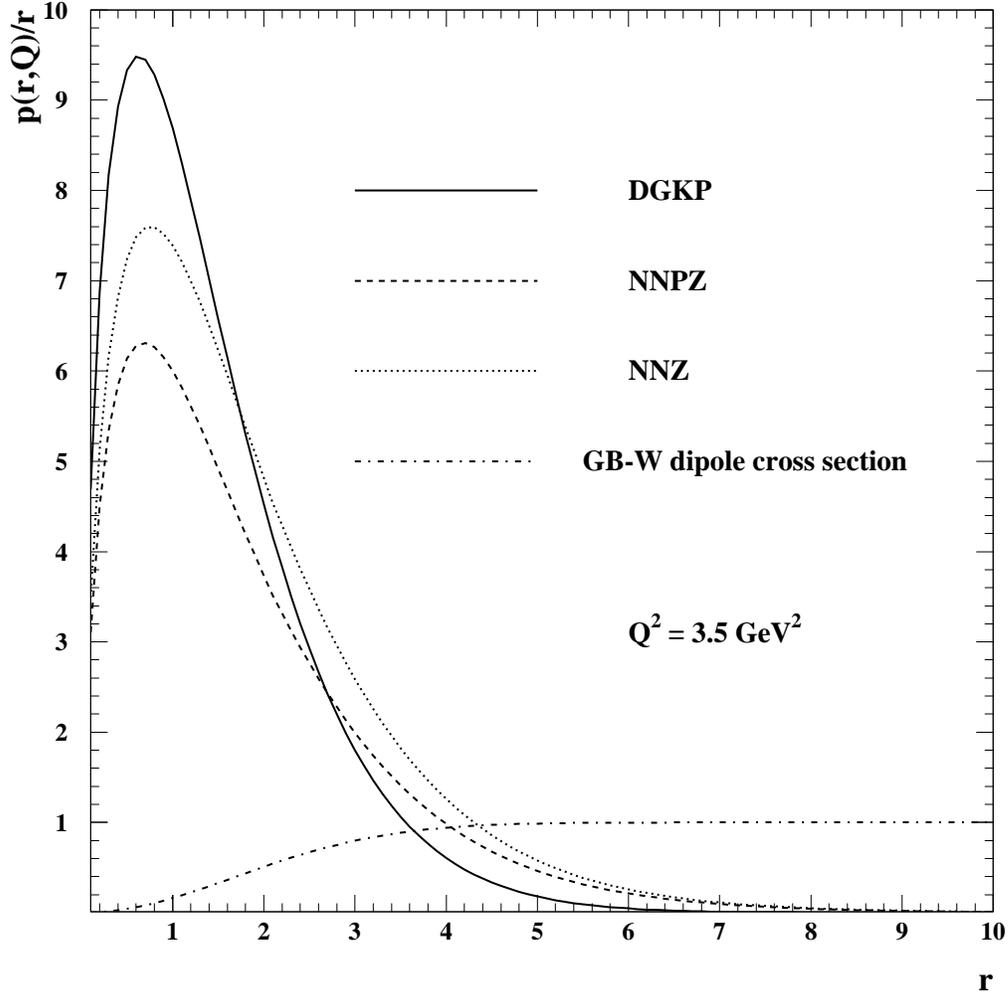,width=15cm}
\end{center}
\caption{The overlap of the wave functions of the photon and vector meson
$p(r,Q)$ plotted as a function of $r$. 
Solid curve: DGKP \cite{DOSCH}, dashed curve NNPZ(1997) \cite{NIKO1}, 
dotted curve NNZ(1994) \cite{NIKO2}. The dashed-dotted curve corresponds to the dipole
cross section by Golec-Biernat and W\"usthoff \cite{GBW} calculated
for $x=10^{-3}$. The relative 
normalization of the dipole cross section and the wave functions is not conserved.}
\label{fig:integrand}   
\end{figure}

In order to understand this discrepancy between the different models, 
we plot in Fig.~\ref{fig:integrand} the 
integrand $p(r,Q)/r $ which appears in Eq.~(\ref{eq:intdistrib})
together with the dipole cross section from the Golec-Biernat and W\"usthoff model.
One can see that different models for the wave functions result
in different shapes in $r$ for $p(r,Q)/r$.
It is clear from this plot that the exact shape of the function $p(r,Q)/r$
for large values of $r$ is very important in this case since it is weighted
by the dipole cross section $\sigma(x,r)$ which is large in this regime.


\subsection{Overlap integral $N(Q)$, mean radius $r_Q$ and theoretical
uncertainty on $S(x,r_Q,b)$}

The results on the forward production cross section 
presented in Fig.~\ref{fig:total} suggest that it is essential
to study in detail the  model dependence of the $S$-matrix  given
by formulae~(\ref{eq:means}) and (\ref{eq:rq}). \\

Let us first study the properties of the quantity $N(Q)$. 
In Fig.~{\ref{fig:limite}} we plot
it for the three different models for the vector meson wave 
function.  In the region of interest, $0.05 < Q^2 < 10.0\, \rm GeV^2$
they result in comparable $N(Q)$ within $10-25 \%$. 
This means that the average $S$-matrix obtained
using Eq.(\ref{eq:means}) is quite robust to a given choice of the 
model for the vector meson wave function.
In fact one can give some very general bounds on $N(Q)$,
nearly independent on the model for $\psi_V$.
From these upper bounds on $N(Q)$ we could obtain an absolute upper bound
for the $S$-matrix which would be model-independent.
We refer to appendix B for technical details.
We represent these bounds in Fig.~\ref{fig:limite}. One can see that
the models happen to be quite close to these limits (within say $25\%$).\\ 

Second, we estimate 
the variation of  $r_Q$ with respect to the different
models for the wave functions. We compute the mean value $r_Q$ as: 
\begin{equation}
r_Q = e^{\langle \log r \rangle_p} \, .
\label{eq:myrq}
\end{equation}
The results for the different models at different values of
$Q^2$ are given in Tab.~\ref{tab:tab1}.  
The values of $r_Q$ are consistent within $25\%$ for the various models.
This confirms the validity of formula~(\ref{eq:rq}). \\

\begin{table}
\begin{center}
\begin{tabular}{|c||c|c||c|c||c|c|}
\hline
 & \multicolumn{2}{|c||}{DGKP} & \multicolumn{2}{|c||}{NNZ} 
 & \multicolumn{2}{|c|}{NNPZ}\\
\hline
$Q^2 \, \rm [GeV^2]$  & $ r_Q\, \rm [fm]$ & $A$
  & $ r_Q\, \rm [fm]$ & $A$ 
  & $ r_Q\, \rm [fm]$ & $A$ 
\\
\hline\hline
0.45 & 0.35 & 1.88 & 0.49 & 2.63 & 0.52  & 2.79 \\
\hline
3.5 & 0.21 & 2.24 & 0.26 & 2.77 & 0.26 &  2.77 \\
\hline
7 & 0.16 & 2.32 & 0.20 & 2.90 & 0.19 &  2.75 \\
\hline
27 & 0.09 & 2.49 & 0.11 & 3.04 & 0.10 & 2.76 \\
\hline
\end{tabular}
\end{center}
\caption{{\em
The mean values $r_Q$ of the dipole size for different 
$Q^2$ obtained in models \DGKP~\cite{DOSCH}, \NNZ~\cite{NIKO2} and \NNPZ~\cite{NIKO1}.
$r_Q$ is evaluated according to formula (\ref{eq:myrq}).
Consistency with formula (\ref{eq:rq}) is checked by quoting 
the corresponding values of $A$, all quite close to 2 and hardly
dependent on $Q^2$.} 
}
\label{tab:tab1}
\end{table}

The three models appear to be more or less equivalent for our purpose.
For simplicity, we then only consider model DGKP~\cite{DOSCH} in the following.


\section{Impact parameter analysis of HERA data}
\label{sec:3}

In this section we extract the $b$-dependent $S$-matrix from
the HERA data.
To this aim, we apply formulae (\ref{eq:means},\ref{eq:rq}) to
the data, 
together with expressions (\ref{eq:doschp},\ref{eq:nikop}) obtained from
the discussion of the vector meson wave function in the previous section.
We also try to deduce from these results the value of the saturation scale.

\subsection{Profile $S(b)$}

The experimental data 
for diffractive production of vector mesons
are usually parametrized by the forward diffractive cross section 
$d\sigma/dt|_{t=0}$ and the logarithmic
slope in momentum transfer $B(t)$ as follows:
\begin{equation}
\frac{d\sigma}{dt}=\frac{d\sigma}{dt}_{|_{t=0}}\cdot e^{-B(t)\cdot t} \, .
\end{equation}
We take the available data \cite{ZEUS,H1} for the electroproduction of
longitudinally polarized $\rho_0$ vector mesons.
These data are given for momentum transfer $t$ 
below $t_{\max}=0.6\,\mbox{GeV}^2$, which
enables us to determine reliably the $S$-matrix only for impact
parameter values larger than $b=1/\sqrt{t_{\max}}\simeq 0.3\,\mbox{fm}$.
In order to compute the Fourier transform appearing in
Eq.~(\ref{eq:means}), we have to assume an extrapolation of
$d\sigma/dt$ to larger
values of $t$.
The formula we use is
\begin{multline}
S(x,r_Q,b)=1-\frac{1}{2\sqrt{\pi}N(Q)}
\sqrt{\frac{d\sigma}{dt}}_{|_{t=0}}
\bigg(
\int_0^{t_{\max}}
dt\,J_0(b\sqrt{t}) e^{-B(t)\cdot
t/2}\\
+\int_{t_{\max}}^{+\infty}dt\,J_0(b\sqrt{t})\sqrt{{\mathcal
E}(t)}\bigg)\ .
\label{eq:sexp}
\end{multline}
Assuming that $B(t)$ is non-increasing with $t$, as various
sets of data seem to indicate,
we choose different functional forms for the extrapolating functions
${\mathcal E}(t)$:
\begin{itemize}
\item an exponential form $\exp(-d\cdot t)$, with the constant $d$
being of the order of $B(t_{\max})$. This choice provides an
upper bound on the $S$-matrix for $b$ close to 0.
\item a power law form $t^{-\alpha}$. This choice is motivated by the fact
that the data for photoproduction at high $t$ \cite{ZEUS2} indicate
a $t$-dependence governed by such a power law, with $\alpha=3$. 
Such less steep $t$-dependence was also obtained in a theoretical
calculation \cite{GINZBURG}.
We consider here two values for $\alpha$: $\alpha=3$ and $\alpha=6$.
\end{itemize}
The parameters of these functions ${\mathcal E}(t)$ are fixed 
from the fit to the
experimental points corresponding to the highest values for $t$,
in the region $0.4\,\mbox{GeV}^2<t<0.6\,\mbox{GeV}^2$.
These specific assumptions give an idea of the uncertainty on the
determination of $S(x,r_Q,b)$ due to our lack of knowledge of the
differential cross section for $t>t_{\max}$.
Another source of uncertainty is the experimental errors on the
measurements of $B(t)$ and $d\sigma/dt|_{t=0}$.
A complete error analysis would lie beyond the scope of this paper.
Here we just estimate the influence of these uncertainties on $S(x,r_Q,b)$ 
by varying the measured quantities inside
the 1-$\sigma$ error bars.
By this method, we believe that we obtain a strict overestimate of the errors.
The theoretical input $N(Q)$ neeeded in formula (\ref{eq:sexp}) is
computed within the model \DGKP. 
The value $r_Q$ is estimated using the procedure described in
Sec.~(\ref{sec:1}) and~(\ref{sec:2}). The other models
\NNPZ~and \NNZ~have also been tested and lead to very similiar results.\\

We take three values of $Q^2$: $Q_1^2=0.45\,\mbox{GeV}^2$,
$Q_2^2=3.5\,\mbox{GeV}^2$ and $Q_3^2=7\,\mbox{GeV}^2$. These values
correspond to bins of the ZEUS analysis \cite{ZEUS}, which we
consider in the following. We note that the H1 data \cite{H1} lead to similar results.
We recall that these values of $Q^2$ correspond to the respective
values of the dipole size $r_Q$ (see Tab.~(\ref{tab:tab1})):
${r_Q}_1=0.35\;\mbox{fm}$, ${r_Q}_2=0.21\;\mbox{fm}$,
${r_Q}_3=0.16\;\mbox{fm}$.
For these values of $Q^2$, we take similar low values of $x$, namely
$x_1=4.7\cdot 10^{-4}$, $x_2=4.3\cdot 10^{-4}$ and $x_3=5.8\cdot 10^{-4}$
respectively.
The experimental slope $B(t)$ is parametrized
in the region $t<0.6\,\mbox{GeV}^2$ 
by $B(t)\!=\!B_0\!-\!ct$, where the values of $B_0$ and $c$ are
$B_0=9.5\, {\rm GeV^{-2}}$,
$c=4.0\, {\rm GeV^{-4}} ({\rm for} \, Q^2 = 0.45\, {\rm GeV^2})$  
and $c=6.1\, {\rm GeV^{-4}} 
({\rm for } \, Q^2=3.5\, {\rm GeV^2}\ {\rm and}\  Q^2=7\, {\rm GeV^2}$.\\

\begin{figure}[ht]
\begin{center}
\epsfig{file=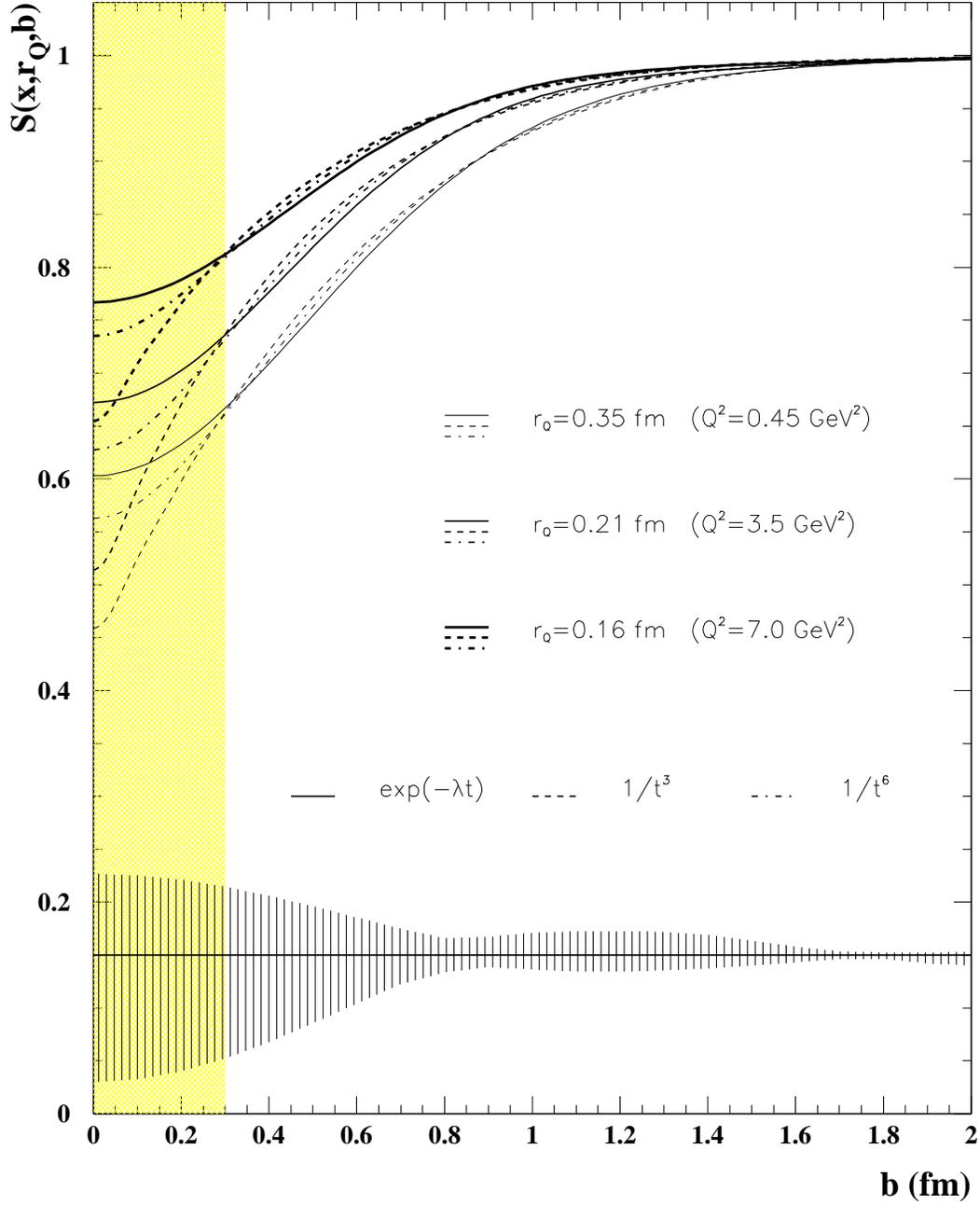,width=15cm}
\end{center}
\caption{$S$-matrix for dipole-proton scattering as a function of
impact parameter $b$.
Three different $Q^2$ are considered, corresponding to three
typical values for the size of the interacting dipole (which are
estimated according to $r_Q\equiv e^{<\log r>_p}$).
For each value of $Q^2$, the curves corresponding to three extrapolations of the data for
$t>0.6\;\mbox{GeV}^2$ are shown. The shaded band indicates the region
of impact parameter $b$ where the choice of this extrapolation is
crucial, and thus where our extraction is not reliable.
The hashed band on the bottom is an estimate of the errors due to the
experimental uncertainties on $d\sigma/dt|_{t=0}$ and on $B(t)$ 
(for $t<0.6\,\mbox{GeV}^2$).
}
\label{fig:results}
\end{figure}

The results for $S(x,r_Q,b)$ are shown in Fig.~(\ref{fig:results}),
and are obtained using Eq.~(\ref{eq:sexp}) applied to the ZEUS data.
For each $Q^2$, the curves corresponding to three choices for the extrapolation
function ${\mathcal E}(t)$ are drawn. The uncertainty corresponding to the
errors on the measurements themselves are computed for the
intermediate value of $Q^2$, and are represented in the lower
part of the plot by a hashed band.

First, we observe that for $b>0.3\,\mbox{fm}$, the curves corresponding to
different values of $Q^2$ are clearly separated and ordered according to:
\begin{equation}
S(x_1,r_{Q_1},b)<S(x_2,r_{Q_2},b)<S(x_3,r_{Q_3},b)\ .
\end{equation}
As $x_1\simeq x_2\simeq x_3$, this means that
the proton is less transparent to dipoles of larger size $r$.
We also note that in this region of $b>0.3\;\mbox{fm}$, the results are
not very dependent on the choice of extrapolation of the
$t$-dependence of the data.
However, for $b<0.3\;\mbox{fm}$ (shaded region in
Fig.~(\ref{fig:results})), 
the results become
very sensitive to the asymptotic behaviour of the cross section
$d\sigma/dt$, and data at larger $t$ are needed to be able to make
firm statements about the value of $S(x,r_Q,b)$ near $b=0$.
We also note that the errors on the measurements for
$t<0.6\,\mbox{GeV}^2$ lead at most to an uncertainty of about 
$25 \,\%$ for the $S$-matrix at impact parameter $b=0.3\, \rm fm$.\\

Finally, 
one can compute the total cross section for dipole-proton scattering using
the following formula (which is nothing more than Eq.~(\ref{eq:landau})
averaged over $r$, for real $S$-matrix):
\begin{equation}
\langle \sigma_{\mbox{\footnotesize tot}}^{q\bar q-p}(x,r)\rangle_p
=2\int d^2{\mathbf b}\,
(1-\langle S(x,r,b)\rangle_p)\ .
\end{equation}
Using formula (\ref{eq:means}) to replace $\langle S(x,r,b)\rangle_p$ 
and performing the integral over $\mathbf b$, one
sees that this cross section is determined by $N(Q)$ and
by the forward differential cross section 
$d\sigma/dt_{|t=0}$.
For photon virtualities $Q_1^2$, $Q_2^2$ and $Q_3^2$, one obtains the
values $\langle\sigma_{\mbox{\footnotesize tot}}^{q\bar q-p}\rangle_p
 =14.9\,\rm mb$, $10.6\,\rm mb$ and $7.5\,\rm mb$ respectively.
These results are consistent with those of 
Ref.~\cite{GBW,NIKO1,MFGS,RUETER,SHAW,CVETIC}, however, the
relationship between $Q^2$ and the radius $r_Q$ quoted in these papers
remains a large source of uncertainty in this comparison.


\subsection{Towards an estimate of the saturation scale}

One can try to estimate the quark saturation scale $Q_s^2$ using the results
presented in Fig.~\ref{fig:results}. From this analysis it is in principle possible
to extract the dependence of this scale on the impact parameter $b$
of the collision. We try the following phenomenological formula \cite{MUELLERS}
for the $S$-matrix:
\begin{equation}
S(x,r_Q,b) \, = \, \exp(-r_Q^2 Q_s^2(x,b) / 4) \,\ .
\label{eq:sexpqs}
\end{equation}
This exponential form comes from a Glauber-like summation of multiple
independent scatterings of the quark-antiquark pair on the target
nucleon \cite{M,AGL}.
It is also supported by the Golec-Biernat and W\"usthoff model \cite{GBW}
where the radius $R_0(x)$ introduced there can be related to the
saturation scale by $R_0(x) = 1 / Q_s(x,b)$.
However, in that work,
it was assumed that the $b$-profile of the $S$-matrix has the form of
a sharp cutoff at a distance of order 1 fm.

Using the approximation $r_Q \simeq 2/\sqrt{Q^2 + m_V^2}$ one obtains 
$Q_s^2 \simeq 1-1.5 \,\rm GeV^2$ and  $Q_s^2 \simeq 0.2\, \rm GeV^2$
for $b=0.3 \, \rm fm$ and $b=1.0 \, \rm fm$ respectively.
These values are consistent in order of magnitude when computed at different
$Q^2$, which confirms the relevance of formula~(\ref{eq:sexpqs}).
These results would suggest that at small impact parameter values the saturation
scale is in the semi-hard regime and support the onset of the shadowing
effects at HERA. One should however stress that these estimates are rather
rough and should be taken with care since they strongly depend on the
precise value of $r_Q$.
Indeed, $r_Q$ comes squared in the formula for $Q_s^2$.


\section{Summary}
\label{sec:4}
In this paper we have shown that using the HERA data on diffractive 
electroproduction of vector mesons it is possible to extract the $S$-matrix
element $S(x,r,b)$ for dipole-proton scattering in impact parameter space.
By considering the full $t$ dependence of this process 
we have shown how to obtain the $S$-matrix element averaged over
dipole sizes $r$. Due to the particular properties of the 
overlap function between photon and meson wave functions  this
average can be replaced by the $S$-matrix element evaluated at 
$r\!=\!r_Q \!\sim\!2/\sqrt{Q^2\!+\!m_V^2}$. 
We have then shown that our results
on the $S$-matrix are only weakly dependent on the choice for
the model of the meson wave function.
Since the data are available only for low momentum transfer, $t<0.6\,\rm GeV^2$, 
we have shown that our results are reliable for impact parameter
values larger than $b \simeq 0.3 \, \rm fm$. It would be very interesting
to extend the measurement of diffractive electroproduction of vector mesons
to larger values of $t$. This would enable us to explore the very interesting
regime of central impact parameter collisions at HERA. 
We have also estimated
the value of the dipole cross section integrated over the impact parameter 
and have found it to be consistent with other analyses. 
Finally we have discussed how to use
this result to estimate the saturation scale at HERA collider.
We have suggested that  the onset of the shadowing effects can be dependent
on the impact parameter of the collision and that for central collisions
it could occur in the semihard regime.  
However further theoretical and experimental studies are necessary.

\section*{Acknowledgements}
We thank Marcello Ciafaloni for his comments,
and Allen C. Caldwell and Mara S. Soares for correspondance. 
S.M. thanks Columbia University for welcome at a
preliminary stage of this work, and the Service de Physique
Th\'eorique de Saclay for support at that time. He also thanks Robi Peschanski
and Bernard Pire for their suggestions. 
A.M.S. thanks Krzysztof Golec-Biernat and Jan Kwieci\'nski for interesting discussions.
A.H.M. wishes to thank Marcello Ciafaloni for his
hospitality at the University of Florence where this collaboration
began, and he wishes to thank Dominique Schiff for her hospitality
in Orsay.


\bibliography{mesons}

\begin{thebibliography}{999}

\bibitem{GBW}  K. Golec-Biernat, M. W\"usthoff,  
             {\it Phys. Rev.} {\bf D59}  (1999) 014017;
             {\it Phys. Rev.} {\bf D60} (1999)  114023.  

\bibitem{AMALDI} U. Amaldi, K.R. Schubert,
{\it Nucl.\ Phys. } {\bf B166} (1980) 301.

\bibitem{DIPOLENIK}    N.N. Nikolaev and B.G. Zakharov, {\em Z. Phys.} {\bf C49}   
(1991) 607; {\em Z. Phys} {\bf C53} (1992) 331; {\em Z. Phys.} {\bf C64}   
(1994) 651; {\em JETP} {\bf 78} (1994) 598.    

\bibitem{DIPOLEMUELL}  A. H. Mueller, {\em Nucl.\ Phys.} {\bf B415} (1994) 373;
   A. H. Mueller and B. Patel, {\em Nucl.\ Phys.} {\bf B425} (1994) 471;   
A. H. Mueller, {\em Nucl. Phys.} {\bf B437} (1995) 107.  

\bibitem{Bjorken:1971ah}
J.~D.~Bjorken, J.~B.~Kogut and D.~E.~Soper,
{\it Phys.\ Rev.} {\bf D3} (1971) 1382.

\bibitem{MUELLER} A.H. Mueller, {\it Eur.\ Phys.\ J.} {\bf A1} (1998)
19.

\bibitem{landau} L.D. Landau, E.M. Lifshitz, {\it Quantum Mechanics},
Mir, 1966.


\bibitem{DOSCH} H.G. Dosch, T. Gousset, G. Kulzinger and H.J. Pirner,
{\it Phys. Rev.} {\bf D55} (1997) 2602.\\
G.~Kulzinger, H.~G.~Dosch and H.~J.~Pirner,
{\it Eur.\ Phys.\ J.\ } {\bf C7} (1999) 73.

\bibitem{NIKO2} J. Nemchik, N.N. Nikolaev, B.G. Zakharov ,
 {\it Phys.Lett.} {\bf B341} (1994) 228.

\bibitem{NIKO1}J. Nemchik, N.N. Nikolaev, E. Predazzi, B.G. Zakharov, 
{\it Z.Phys.} {\bf C75} (1997) 71. 



\bibitem{MFGS} M. McDermott, L. Frankfurt, V. Guzey, M. Strikman, 
{\it Eur. Phys. J. } {\bf C16} (2000) 641.


\bibitem{RUETER}
        M.~Rueter, H.G.~Dosch, {\it Phys. Rev. } {\bf D57} (1998) 4097.


\bibitem{SHAW} J .R. Forshaw, G. Kerley, G. Shaw,  
{\it Phys. Rev. } {\bf D60} (1999) 074012.

\bibitem{CVETIC} G. Cvetic , D. Schildknecht, A. Shoshi,
{\it Acta Phys. Polon. } {\bf B30} (1999) 3265.

\bibitem{CAPELLA} A. Capella, E.G. Ferreiro, C.A. Salgado, A.B. Kaidalov,
{\it Nucl. Phys. } {\bf B593} (2001)  336; {\it Phys.\ Rev.\ } {\bf D63}
(2001) 054010.


\bibitem{BRODSKY}
 S.J.~Brodsky, L.~Frankfurt, J.F.~Gunion, A.H.~Mueller, M.~Strikman,{\it
Phys. Rev. } {\bf D50} (1994) 3134.

\bibitem{FKS}
 L.~Frankfurt, W.~Koepf, M.~Strikman,
{\it Phys. Rev. } {\bf D54} (1996) 3194.


\bibitem{CALDWELL} A. Caldwell and M.S. Soares, {\tt hep-ph/0101085}.


\bibitem{ZEUS}   ZEUS collaboration, {\it Eur. Phys. J.} {\bf C6} (1999) 603.

\bibitem{H1} H1 collaboration {\it Eur. Phys. J.} {\bf C13} (2000) 371. 

\bibitem{ZEUS2}  ZEUS collaboration, {\it Study of the diffractive
production of vector mesons at large $Q^2$ or at large $|t|$ at HERA},
EPS 1999, Tampere.

\bibitem{GINZBURG}
    D.Yu.~Ivanov, {\it Phys. Rev. } {\bf D53} (1996) 3564;
    I.F.~Ginzburg, D.Yu.~Ivanov,  {\it Phys. Rev. } {\bf D54} (1996) 5523.



\bibitem{MUELLERS}  A.H. Mueller, Lectures given at International
Summer School on Particle Production Spanning MeV and TeV Energies
(Nijmegen 99), 
Nijmegen, Netherlands, 8-20 Aug 1999,
and at MEETING-NOTE = 17th Autumn School: QCD: Perturbative or
Nonperturbative? (AUTUMN 99), Lisbon, Portugal, 29 Sep - 4 Oct 1999. 
{\tt hep-ph/9911289}.


\bibitem{M} A.H. Mueller, {\it Nucl. Phys.} {\bf B335} (1990) 115.

\bibitem{AGL} A.L. Ayala Filho, M.B. Gay Ducati, E.M. Levin, {\it
Nucl. Phys.} {\bf B493} (1997) 305; {\it Nucl. Phys.} {\bf B511}
(1998) 355.

\end{thebibliography}

\clearpage

\appendix

\section{Models for the meson wave function}

\label{sec:models}

In this appendix, we detail the phenomenological forms we adopt for the meson
wave function.

\subsection{Model by Dosch, Gousset, Kulzinger, Pirner
(DGKP, Ref.~\cite{DOSCH})}

The form of the vector meson wave function is given by:
\begin{equation}
\psi^{h,\bar h}_{V}(z,r)=\delta_{h,-\bar h}\frac{\calN}{\sqrt{4\pi}}
4\omega(z(1\!-\!z))^{3/2}\exp\left(-\frac12\frac{m_V^2}{\omega^2}(z\!-\!1/2)^2\right)
\exp\left(-\frac12\omega^2 r^2\right)\ .
\end{equation}
The parameters chosen are the following:
\begin{equation}
\omega=0.330\;\mbox{GeV} \quad\mbox{and}\quad \calN=4.48\ .
\end{equation}
An additional feature of this model is that 
the mass $m_q$ of the quarks that appears in the photon wave function 
is running with $Q^2$. This gives important effects only for very low $Q^2$ and
for photoproduction. In this regime, it is argued that the quarks should have constituent
mass. The formula for the running mass reads:
\begin{equation}
m_q(Q^2)=\left\{
\begin{split}
0.220\mbox{\,GeV}\cdot(1\!-\!Q^2/Q_0^2)\quad & \mbox{for}\quad Q^2<Q_0^2=1.05\mbox{\,GeV}^2\\
0                                      \quad & \mbox{for}\quad Q^2>Q_0^2
\end{split}\right.
\end{equation}
This model is referred to as \DGKP.

\subsection{Models by Nemchik, Nikolaev, Predazzi, Zakharov
(NNZ(1994), Ref.~\cite{NIKO2}
and NNPZ (1997), Ref.~\cite{NIKO1})}

In this model, the radial wave function $\phi(r,z)$ 
appearing in eq.(\ref{eq:nikop}) satisfies the following normalization 
condition:
\begin{equation}
1=\frac{N_c}{2\pi}\int_0^1\frac{dz}{z^2(1-z)^2}\int d^2{\mathbf
r}\bigg\{
m_q^2\phi^2(z,{\mathrm r})+(z^2+(1-z)^2)(\partial_r\phi(z,{\mathrm
r}))^2\bigg\}\ .
\label{eq:norm2}
\end{equation} 
This equation is nothing else but the normalization condition~(\ref{eq:norm}),
applied to the transversely polarized meson wave function.
The obtained normalization factor $\Psi_0(1S)$ is assumed to be the same for a
longitudinally polarized meson.
The function $\phi(r,z)$ is defined by:
\begin{multline}
\phi(z,r)=\Psi_0(1S)\\
\times\bigg\{4z(1-z)\sqrt{2\pi R^2}
\exp\left(-\frac{m_q^2 R^2}{8z(1\!-\!z)}\right)
\exp\left(-\frac{2z(1\!-\!z)r^2}{R^2}\right)\exp\left(\frac{m_q^2 R^2}{2}\right)\\
+C^4\frac{16 a^3(r)}{A(z,r)B(z,r)^3} rK_1(r\cdot
A(z,r)/B(z,r))\bigg\}\ .
\end{multline}
The functions $A$ and $B$ are given by
\begin{equation}
\begin{split}
A^2(z,r)&=1+\frac{C^2a^2(r)m_q^2}{z(1\!-\!z)}-4C^2a^2(r)m_q^2\\
B^2(z,r)&=\frac{C^2a^2(r)}{z(1\!-\!z)}\ ,
\end{split}
\end{equation}
and
$a(r)\equiv3/(8m_q\as(r))$. The
following prescription is taken for the running coupling $\as(r)$:
\begin{equation}
\as(r)=\alpha_0\quad \mbox{for}\quad r>r_s\quad\mbox{and}\quad
\as(r)=\frac{4\pi}{\beta_0\log({1}/{\Lambda^2
r^2})}\quad\mbox{for}\quad r<r_s\ ,
\end{equation}
where the parameter values are
\begin{equation}
r_s=0.42\;\mbox{fm},\quad\alpha_0=0.8,\quad\Lambda=200\;\mbox{MeV}\ .
\end{equation}
The masses of the quarks are taken to be $0.15\;\mbox{GeV}$.

The other parameters $\Psi_0(1S)$, $C$ and $R^2$ are chosen
taking into account several constraints: the normalization condition (\ref{eq:norm2})
for the wave function has to be satisfied, and the value of the
leptonic decay width must agree with the experimental measurement.
Additionally, the mean radius of the meson has to be of the order of a hadronic scale.
Two different sets of parameters are considered:
\begin{equation}
\begin{split}
C&=0.25,\quad R^2=0.76\;\mbox{fm}^2\quad\mbox{(reference \cite{NIKO2})}\\
\mbox{and}\quad
C&=0.36,\quad R^2=1.37\;\mbox{fm}^2\quad\mbox{(reference
\cite{NIKO1})}\ .
\end{split}
\end{equation}
These two choices are refered to as \NNZ~and \NNPZ respectively.


\section{Model-independent bounds on the overlap integral $N(Q)$}
\label{sec:bounds}

In this appendix we explore more the properties of
$N(Q)$ and show that upper bounds can be given on this quantity,
regardless the model adopted for $\psi_V$. These provide upper bounds on $S(x,r_Q,b)$.\\


$N(Q)$ can be seen as a scalar product of the two
wave functions, $N(Q)\!=\!\langle \psi_{\gamma^*} | \psi_{V} \rangle $, 
see Eqs. (\ref{eq:distrib}) and (\ref{eq:intdistrib}).
The Cauchy-Schwartz inequality then applies. It leads to an
upper estimate for the integral: the product of the two
wave functions is smaller than the square root of
the product of the integrals of the squared wave functions. Using the
normalization condition~(\ref{eq:norm}) for the meson wave
function\footnote{
Strictly speaking, this bound is only valid for models for which the
condition~(\ref{eq:norm}) is enforced for the longitudinally polarized
vector meson. 
}
and computing the full integral of the squared photon wave function,
one eventually obtains:
\begin{equation}
N(Q)\leq 2\hat e_V \frac{\aem N_c}{6\pi}
\left(1-6\frac{m_q^2}{Q^2}+24\frac{m_q^4}{Q^4}\frac{1}{\sqrt{1\!+\!\frac{4m_q^2}{Q^2}}}
\tanh^{-1}\frac{1}{\sqrt{1\!+\!\frac{4m_q^2}{Q^2}}}\right)\ .
\label{eq:bound1}
\end{equation}
The r.h.s. of eq.(\ref{eq:bound1}) grows with $Q^2$ and thus this
inequality is only interesting in the small-$Q^2$ region.
Although independent of the precise form of the vector meson wave function, 
this bound nevertheless depends on the masses of the quarks.\\

Second, with the additional assumption that $\psi_V(z,{\mathbf r})$ is maximum for
$r=0$, we can write:
\begin{equation}
N(Q)\leq 2\hat e_V\int d^2{\mathbf r}\, dz\,\psi_{\gamma^*}(z\!=\!1/2,{\mathbf r};Q)\,
\psi_V(z,{\mathbf r}\!=\!0)\ .
\end{equation}
The integrals over $z$ and ${\mathbf r}$ in the r.h.s. are now factorized. The
integration over $\mathbf r$ is performed analytically, while the
one over $z$, involving $\psi_V$, can be expressed as a function of
the coupling $f_V$ of the meson to the electromagnetic current, using
the relation~(\ref{eq:lwid}).
This finally leads to:
\begin{equation}
N(Q)\leq 2\sqrt{\pi\aem}f_V\frac{Q}{Q^2+4m_q^2}\ .
\end{equation}
The r.h.s. vanishes like $1/Q$ at large Q, which makes this inequality
useful for large $Q$.

\begin{figure}[ht]
\begin{center}
\epsfig{file=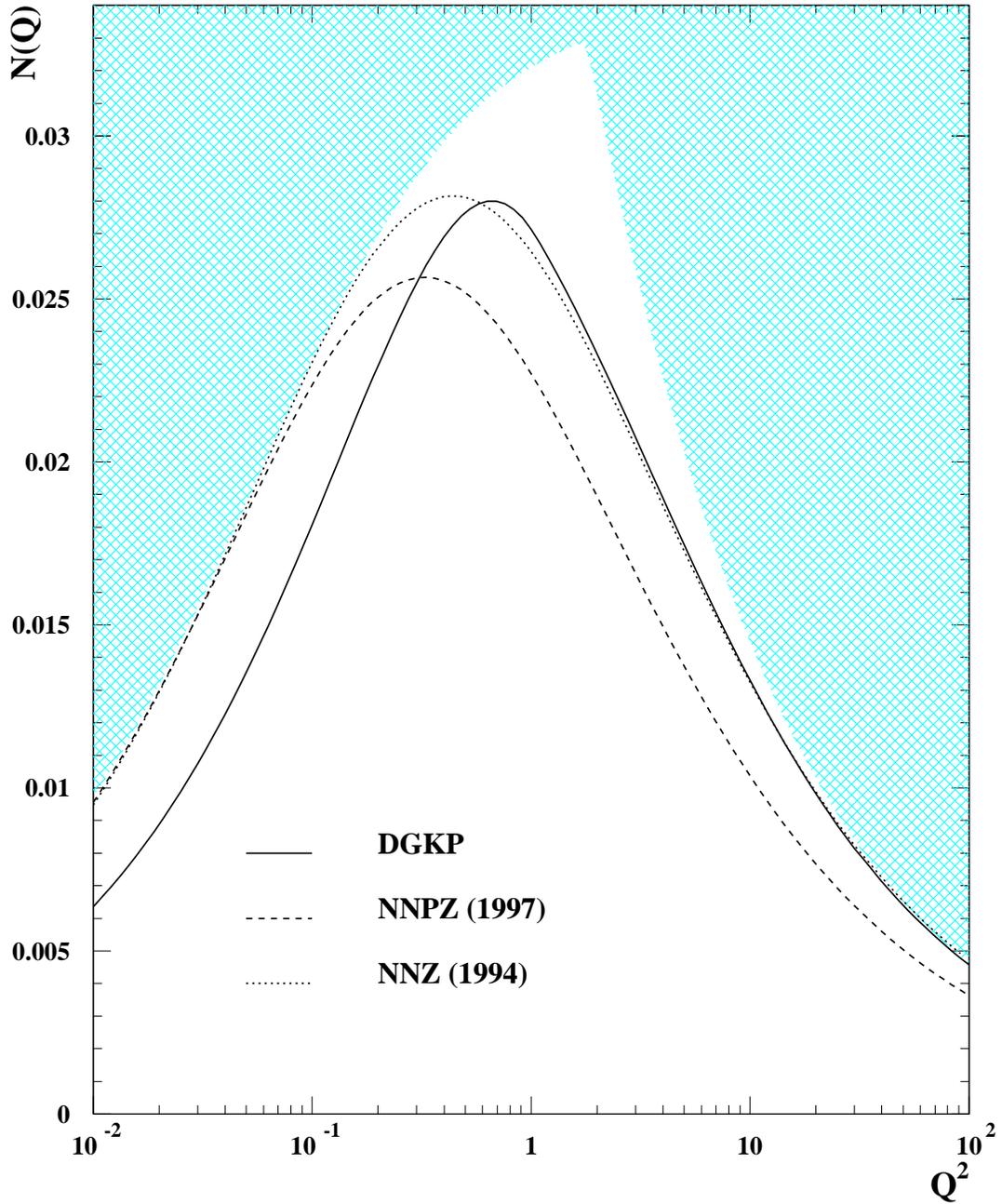,width=15cm}
\end{center}
\caption{Integrated dipole distribution $N(Q)$ as a function of $Q^2$.
The curves corresponding to the
different models for the meson wave function are shown. 
The shaded area is theoretically forbidden.}
\label{fig:limite}
\end{figure}

The region in the ($Q^2$,$N(Q)$) plane which is forbidden by these
bounds is depicted in Fig.~(\ref{fig:limite}).
One sees that all the models are very close to the upper
bounds. 

\clearpage

\clearpage
\newpage
\clearpage
\clearpage

\newpage

\end{document}